\documentclass[utf8]{FrontiersinVancouver} 

\usepackage{url,hyperref,lineno,microtype,subcaption}
\usepackage[onehalfspacing]{setspace}
\usepackage{amsmath,amssymb,amsfonts}
\usepackage{algorithmic}

\usepackage{enumitem}
\setlist[itemize]{noitemsep, topsep=0pt}
\usepackage{graphicx}
\usepackage{textcomp}
\usepackage{xcolor}
\usepackage{array}
\usepackage{caption}
\usepackage{booktabs,multirow}
\usepackage{float}
\usepackage{geometry}
\usepackage{xr}
\usepackage[capitalise]{cleveref}

\def\keyFont{\fontsize{8}{11}\helveticabold }
\def\firstAuthorLast{Kiran and Riedel} 
\def\Authors{Yadu Kiran\,$^{1}$, Marc Riedel\,$^{1}$}
\def\Address{$^{1}$ University of Minnesota, Department of Electrical and Computer Engineering, Minneapolis, Minnesota, United States}
\def\corrAuthor{Yadu Kiran, 200 Union St. S.E., Minneapolis, Minnesota 55455}

\def\corrEmail{kiran013@umn.edu}

\begin{document}
\onecolumn
\firstpage{1}

\title {A Scalable Approach to Performing Multiplication and Matrix Dot-Products in Unary} 

\author[\firstAuthorLast ]{\Authors} 
\address{} 
\correspondance{} 

\extraAuth{}

\maketitle

\begin{abstract}

Stochastic computing is a paradigm in which logical operations are performed on randomly generated bit streams. Complex arithmetic operations can be executed by simple logic circuits, resulting in a much smaller area footprint compared to conventional binary counterparts. However, the random or pseudorandom sources required for generating the bit streams are costly in terms of area and offset the advantages. Additionally, due to the inherent randomness, the computation lacks precision, limiting the applicability of this paradigm. Importantly, achieving reasonable accuracy in stochastic computing involves high latency.
Recently, deterministic approaches to stochastic computing have been proposed, demonstrating that randomness is \emph{not} a requirement. By structuring the computation deterministically, exact results can be obtained, and the latency greatly reduced. The bit stream generated adheres to a ``unary'' encoding, retaining the non-positional nature of the bits while discarding the random bit generation of traditional stochastic computing. This deterministic approach overcomes many drawbacks of stochastic computing, although the latency increases quadratically with each level of logic, becoming unmanageable beyond a few levels.
In this paper, we present a method for \emph{approximating} the results of the deterministic method while maintaining low latency at each level. This improvement comes at the cost of additional logic, but we demonstrate that the increase in area scales with $\sqrt{n}$, where $n$ represents the equivalent number of binary bits of precision. Our new approach is general, efficient, composable, and applicable to all arithmetic operations performed with stochastic logic. We show that this approach outperforms other stochastic designs for matrix multiplication (dot-product), which is an integral step in nearly all machine learning algorithms.

\end{abstract}
\section{Introduction}

In stochastic computing, randomly generated streams of 0's and 1's are used to represent fractional numbers. The number represented by a bit stream corresponds to the probability of observing a 1 in the bit-stream at any given point in time. The advantage of this representation is that complex operations can be performed with simple logic, owing to the non-positional nature of the bits. For instance, multiplication can be performed with a single AND gate, and scaled addition can be performed with a single multiplexer. The simplicity and scalability of these operations make computing in this domain very appealing for applications that handle large amounts of data, especially in the wake of Moore's Law slowing down. Machine learning models are one such application that ticks all the boxes.

The drawbacks of the conventional stochastic model are as follows: 1) the latency is high, and 2) due to randomness, the accuracy is low. Latency and accuracy are related parameters: to achieve acceptable accuracy, high latency is required~\cite{armin13}. Recently, a ``deterministic'' approach to stochastic computing has been proposed~\cite{devon2} that uses all the same structures as stochastic logic but on deterministically generated bit streams. Deterministic approaches incur lower area costs since they generate bit streams with counters instead of expensive pseudo-random sources such as linear feedback shift registers (LFSRs). Most importantly, the latency is reduced by a factor of approximately $\frac{1}{2^n}$, where $n$ is the equivalent number of bits of precision. However, the latency is still an issue as it increases quadratically for each level of logic. Any operation involving two $2^n$-bit input bit streams will produce a resulting bit stream of length $2^{2n}$ bits. This is a mathematical requirement: for an operation such as multiplication, the range of values of the product scales with the range of values of the operands. However, most computing systems operate on constant precision operands and products. Since this is not sufficient to represent the $2^{2n}$ output in full precision, we will have approximation errors. Our primary goal is to minimize this error.

Recent papers have discussed techniques for approximating the deterministic computation with quasirandom bit streams, such as Sobol sequences~\cite{riedel1,lfsrs,sobol,sobolckt}. Unfortunately, the area cost of these implementations is high: the logic to generate the quasirandom bit streams is complex and grows quickly as the number of bit streams increases, in most cases completely offsetting the benefits.

In this paper, we present a scalable deterministic approach that maintains constant bit stream lengths and approximates the results. This approach has much lower area cost than the quasirandom sequence approach. We structure the computation by \emph{directly} pairing up corresponding bits from the input bit streams using only simple structures such as counters. Not only does our approach achieve a high degree of accuracy for the given bits of precision, but it also maintains the length of the bit streams. This property lends \emph{composability} to our technique, allowing multiple operations to be chained together. Maintaining a constant bit stream length comes at the cost of additional logic, but we demonstrate that the increase in area scales with $\sqrt{n}$, where $n$ is the number of binary bits of precision. The new approach is general, efficient, and applicable to all arithmetic operations performed with stochastic logic. It outperforms other state-of-the-art stochastic techniques in both accuracy and circuit complexity. We also evaluate our approach with matrix dot-product, an integral set in machine learning algorithms. We demonstrate that our approach is a good fit for machine learning, as it allows one to increase the precision of the inputs while preserving the bit-length/latency at the output.

As the bit streams are no longer random, the term ``stochastic'' would be an oxymoron. The bit streams generated for any particular operand follow a ``unary'' encoding, where all the 1's are clustered together, followed by all the 0's (or vice versa). Hence, we shall refer to this approach as ``unary'' computing in this paper.

This paper is structured as follows: \cref{sec:background} provides a brief overview and background of stochastic computing. \cref{sec:OurApproach} presents our new approach. \cref{sec:Proof} provides the mathematical reasoning behind our design. \cref{sec:HardwareImplementation} details the gate-level implementation. \cref{sec:Results} evaluates our method and compares and contrasts it with prior stochastic approaches. Finally, \cref{sec:conclusion} outlines the implications of this work.

\section{Background Information} \label{sec:background}
\subsection{Introduction to Stochastic Computation} \label{ssec:background_1}

The paradigm of stochastic logic (sometimes called stochastic ``computing'') operates on non-positional representations of numbers~\cite{gaines}. Bit streams represent fractional numbers: a real number $x$ in the unit interval (i.e., $0 \le x \le 1$) corresponds to a bit stream $X(t)$ of length {\it L}, where $t=1,2,...,L$. If the bit stream is randomized, then for precision equivalent to conventional binary with precision $n$, the length of the bit stream {\it L} must be $2^{2n}$\cite{qian11-PhD}. The probability that each bit in the stream is 1 is denoted by $P(X=1)=x$. Below is an illustration of how the value $\frac{5}{8}$ can be represented with bit streams. Note that the representation is not unique, as demonstrated by the four possibilities in the figure. There also exists a bipolar format which can be used to natively represent negative numbers, but for the sake of simplicity, we shall restrict our discussions to the unipolar format. Although, the concepts which we discuss can also be applied to the bipolar format as well. In general, with a stochastic representation, the position of the 1's and 0's do not matter. 


\begin{figure}[h]
\centering
\includegraphics[scale=1.3]{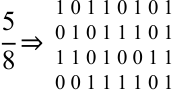}
\label{fig:Stochastic_Representation}
\end{figure}

Common arithmetic operations that operate on probabilities can be mapped efficiently
to logical operations on unary bit-streams.

\begin{figure}[h]
\centering
\includegraphics[scale=1.3]{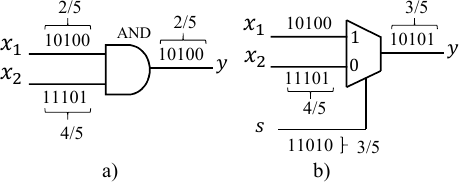}
\caption{}
\label{fig:MultiAdd}
\end{figure}

\begin{itemize}[noitemsep,topsep=0pt]
\item {\bf Multiplication}. Consider a two-input AND gate whose inputs are two independent bit
streams $X_1(t)$ and $X_2(t)$, as shown in
~\cref{fig:MultiAdd}(a). The output bit stream $Y$, is given by
\[
\begin{split}
y &= P(Y=1)= P(X_1 = 1 \textrm{ and } X_2 = 1) \\
&= P(X_1=1) P(X_2=1) = x_1 x_2.
\end{split}
\]
\item {\bf Scaled Addition}. Consider a two-input multiplexer whose inputs are two independent
stochastic bit streams $X_1$ and $X_2$, and its selecting input is a stochastic
bit stream $S$, as shown in ~\cref{fig:MultiAdd}(b). The output bit stream $Y$, is given by
\[
\begin{split}
y &= P(Y = 1) \\ &= P(S = 1) P(X_1 = 1) + P(S = 0) P(X_2 = 1) \\
&= s x_1+ (1 - s) x_2.
\end{split}
\]
\end{itemize}

Complex functions such as exponentiation, absolute value, square roots, and hyperbolic
tangent can each be computed with a small number of  gates~\cite{parhi1,parhi2}.


\subsection{The Deterministic Approach to Stochastic Computing} \label{ssec:recap}

In conventional stochastic logic, the bit streams are generated from a random source such
as a linear feedback shift register (LFSR). The computations performed on these
randomly generated bit streams are not always accurate. The figure below demonstrates a worst-case scenario where multiplying two input
bit-streams corresponding to probabilities $\frac{3}{5}$ and $\frac{2}{5}$, results in an output of
probability $\frac{0}{5}$.

\begin{figure}[h]
\centering
\includegraphics[scale=1.2]{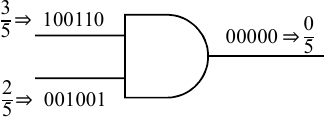}
\label{fig:Conventional_Drawback}
\vspace{-1em}
\end{figure}

Consider instead a {\em unary encoding}, one in
which  all the 1's appear consecutively at the start, followed by all the 0's (or
vice-versa), as shown below. This is also referred by some as ``Thermometer encoding''.

\begin{figure}[h]
\centering
\vspace{-1.0em}
\includegraphics[scale=1.3]{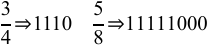}
\label{fig:Thermometer_Encoding}
\vspace{-1em}

\end{figure}

\begin{figure}[h]
\centering
\includegraphics[scale=1.15]{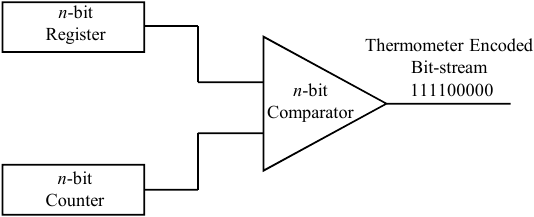}
\caption{}
\label{fig:unary_generator}
\vspace{-1em}
\end{figure}

This encoding is not a requirement, but rather a consequence of the circuit used to generate
deterministic bit streams, shown in ~\cref{fig:unary_generator}. 
For a computation involving $n$-bit precision operands, the
setup involves an $n$-bit register, counter, and comparator.
The register stores the corresponding binary value of the input operand. The
bit stream is generated by comparing the value of the counter to the value stored in the register.
The counter runs from 0 to ${2^n}-1$ sequentially, so the resulting bit-stream
inherits a thermometer encoding.

A ``deterministic'' approach to stochastic computation was proposed, where
the computation is performed on bit-streams which are generated deterministically, resulting in a
unary encoding~\cite{devon2}. By deterministically generating bit streams, all stochastic
operations can be implemented efficiently by maintaining the following property: {\em every bit of one
operand must be matched up against every bit of the other operand(s) exactly once}.

Performing a multiply operation on unary bit-streams using the deterministic approach
involves matching every bit of the first operand, with every bit of the second operand once. This is analogous to a Convolution operation, as illustrated below. Holding a bit of one input operand constant, the operation is repeated
for each of the bits of the other input operand. The particular approach is known as  \emph{clock-division}, due to the division of the clock signal in the circuit for generating the input bit streams.

\begin{figure}[h]
\centering
\includegraphics[scale=1.1]{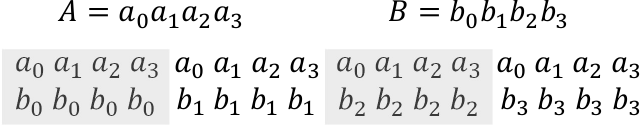}
\label{fig:ClkDiv}
\vspace{-1em}
\end{figure} 

\begin{figure}[H]
\centering
\includegraphics[scale=1.1]{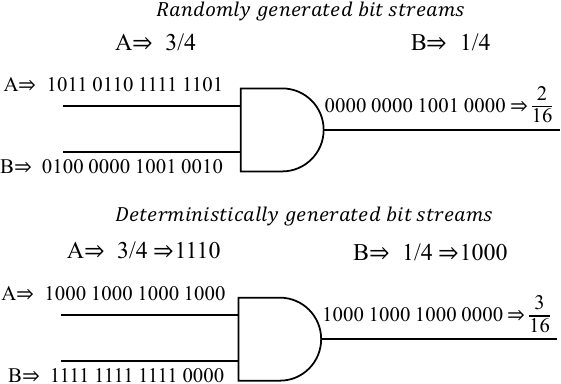}
\caption{}
\label{fig:ClkDiv2}
\vspace{-1em}
\end{figure}

\cref{fig:ClkDiv2} illustrates the Multiply operation on two operands ($\frac{3}{4}$ and $\frac{1}{4}$)
performed stochastically and deterministically. It is evident that the deterministic method achieves perfect accuracy.
However, for each level of logic, the bit stream lengths increase.  For a multiply
operation involving two streams of ${2^n}$ bits each, the output bit stream is ${2^{2n}}$ bits.
This is a mathematical requirement in order to represent the full range of values.
However, for large values of $n$, the bit stream lengths become prohibitive. For most
applications, one has to maintain a constant bit stream length across all the levels of logic, and hence, an approximation is inevitable \cite{fixed-length}. We discuss how to do this in ~\cref{sec:OurApproach}.

\begin{figure}[H]
\centering
\includegraphics[scale=1.2]{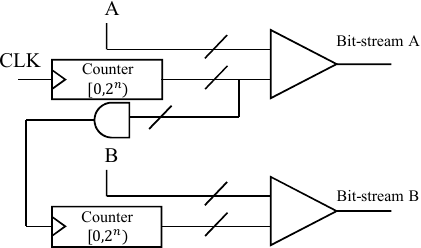}
\caption{}
\label{fig:clock_div_circuit}
\vspace{-1em}
\end{figure}

For an operation such as multiplication, two copies of the circuit in ~\cref{fig:unary_generator}
are used for generating the bit streams of the input operands.  As shown in
~\cref{fig:clock_div_circuit}, the counter of the second input operand counts up
only when the counter of the first input operand rolls over ${2^n}-1$. This can be
achieved by connecting the AND of all the output lines of the first counter to the clock input
of the second counter.

\section {Scalable Deterministic Approach} \label{sec:OurApproach}

In the deterministic approach discussed in Section~\ref{ssec:recap}, the bit stream
lengths grow quadratically with each level of logic \cite{devon2}. This becomes unsustainable for larger circuits. Our goal is to keep the length constant across multiple levels of logic. 

\subsection {Downscaling} \label{ssec:deterministic_downscaling_2}

\begin{figure}[h]
\centering
\includegraphics{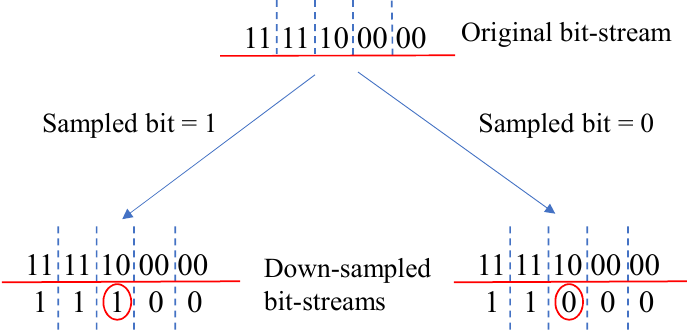}
\label{fig:downscale_2}
\vspace{-1em}
\end{figure}

The low-hanging fruit for approximating is simply to \emph{downscale} the input operands, i.e., generate bit streams of smaller length as shown in \cref{fig:downscale_2}. Consider an input operand that would correspond to a bit stream of length \textit{L}. We want to reduce the length of the generated bit stream by downscaling or approximating the input operand itself. Downscaling is ideally performed by reducing the bit stream by powers of 2, i.e., divide $L$ by $d = 2^i$, where $d$ is the degree of downscaling. In other words, every set of $d$ bits in the original bit stream would correspond to one bit in the downscaled bit stream. The deterministic multiplication operation restores the target length.

Downscaling is easily achieved by right-shifting the value stored in the register in \cref{fig:unary_generator}. For example, for an input operand with $2^4 = 16$ bits of precision and a probability value of $\frac{12}{16}$, we would store the binary equivalent of 12, i.e., $1100_2$, in the register. To downscale the value by a factor of 4, we would right-shift the value of the register by 2 bits to obtain the binary value $11_2$ (which corresponds to the probability value $\frac{3}{4}$). In general, to downscale a value by a factor of $d = 2^i$, we would right-shift by $i$ bits. Consequently, this would also reduce the size of the counters used for bit generation.

In \cref{fig:ClkDiv2}, we showed that deterministically multiplying two input bit streams of length $2^n$ bits each results in an output bit stream of length $2^{2n}$. However, if we were to approximate the input operands to bit streams of length $2^{\frac{n}{2}}$, then our output bit stream would be limited to $2^n$ bits. If the target value of a bit stream can be accurately represented with fewer bits, then there will be no errors. For example, the probability $\frac{20}{32}$ can also be represented as $\frac{10}{16}$ or $\frac{5}{8}$. However, in general, the process of downscaling will introduce errors. We want to minimize the error. In a mathematical sense, we want a scheme that always generates the \emph{optimal approximation}.

In the context of this paper, the {\em error} is the difference between the result
and the optimal approximation, given a target bit stream length.  For example, the
probability $\frac{11}{16}$, when downscaled to 4 bits, can be optimally approximated as
$\frac{3}{4}$ (but not as $\frac{1}{4}$, $\frac{2}{4}$, or $\frac{4}{4}$).

When downscaling a unary encoding, there are only two possible scenarios that can occur, irrespective of the length of the input bit streams. These are illustrated in the figure below, where we try to approximate $\frac{5}{10}$ to be represented with just 5 bits. In both cases, a single bit conveys the wrong information. Using either one of the downscaled bit streams as an input to an arithmetic operation results in an error. The method that we will present in this paper always opts for the right-hand side case, where the downscaled bit stream is an under-approximation of the actual value. The reasoning behind this will be evident in \cref{ssec:DoubleMultiplication}.

For an operation involving two downscaled input operands of ${2^n}$ bits each, it can be mathematically deduced that the error that can occur in the
output bit stream is at most ${(2^n-1)}$ bits out of ${2^{2n}}$ bits. Suppose, for example, we want to multiply two values each with
${(2^4)}$ bits precision (i.e., $\frac{x}{16}$, $\frac{y}{16}$), we could downscale the
operands to ${(2^2)}$ bits precision (e.g., $\frac{p}{4}$, $\frac{q}{4}$), producing an output bit-stream of $16$ bits. The error in
the resulting bit-stream would be restricted to (${2^n-1}$) = 3 bits, out of ($2^{2n}$) =
16 bits. Although this error might seem small, it grows as a function of the bit-stream
length of the inputs as well as the number of logic levels. It's worse than it appears as it grows as a function of the bit-stream length of the inputs, as
well as the number of logic levels. We can do better.

\subsection{Error Compensation} \label{ssec:error_compensation}

The basic idea of our approach is to systematically compensate for the error that we
introduce when down-scaling. We do so during the clock division process.

We illustrate with an example. Consider the multiply operation of two input operands, each of length 16 bits. To restrict the length of the output bit stream to just 16 bits,
we will downscale the input operands that corresponds to a bit stream of 4 bits, a downscaling factor of $\frac{16}{4} = 4$. 
In general, if the input-operands are {\it p bits in length}, we ideally down-scale them to
length {\it q bits}, such that q=$\sqrt{p}$ and that the length of the output bit stream remains the same as the input bit steams.
Consider the case where $A = \frac{5}{16}$ and $B = \frac{15}{16}$ as shown below. Neither of the two
input operands can be downscaled to 4 bits without introducing errors. For each input
operand, we round down, shifting the value stored in the register by 2 bits. 
So $A = \frac{5}{16}$ gets down-scaled to $A' =\frac{1}{4}$, which is equivalent to
$\frac{4}{16}$. $B = \frac{15}{16}$ gets down-scaled to $B' = \frac{3}{4}$, which is
equivalent to $\frac{12}{16}$. We underestimate the value of $A'$ by $\frac{1}{16}$, 
and $B'$ by $\frac{3}{16}$.

\begin{figure}[h]
\centering
\includegraphics[scale=1.4]{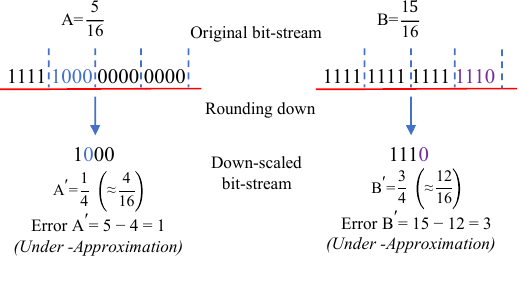}
\vspace{-2 em}
\label{fig:case3a}
\vspace{-1 em}
\end{figure}

Only one bit in a downscaled bit-stream(s) is erroneous. 
And this erroneous bit is carrying \textit{partially incorrect} information. In our example above, 
for $A'$, we can interpret the second bit which is highlighted in blue, as having $1/4$th of its information ``incorrect''.
Likewise, for $B'$, $3/4$th of its last bit (highlighted in orange) can be considered ``incorrect'' information. 

\begin{figure}[h]
\centering
\includegraphics[scale=1.35]{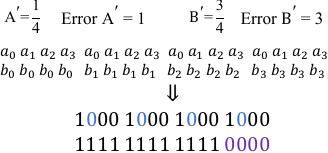}
\label{fig:case3b}
\end{figure}

In normal circumstances, we cannot \textit{correct} a \textit{fractional portion} of a bit; only the bit as a whole. 
However, when performing the clock division operation discussed in \cref{ssec:recap}, each bit is repeated multiple times (in this example,
four times) as shown in the figure above.  This provides the opportunity
to compensate for the error incurred during downscaling. For $A'$, we know that the second bit is erroneous, and that $1/4$th of this bit
is ``incorrect''. This bit is also repeated four times during the clock division operation. So our instinct would be to ``correct'' this error by inverting
that bit once, out of the four times it is repeated. Similarly, for $B'$, we know that $3/4$th of its last bit is ``incorrect''. 
Naturally, we would want to invert this bit three out of the four times it is repeated.

If the input-operands are {\it p bits in length}, we down-scale them to
length {\it q bits}, such that q=$\sqrt{p}$. The down-scaled bit stream has an error of
{\it e}, implying a portion $\frac{e}{q}$, of a 0, is incorrect.  In the clock-division operation, each bit is repeated {\it q}
times. To compensate for the error, we invert the 0 to 1, {\it e} out of the {\it q}
times that it is repeated.

We mentioned earlier in \cref{ssec:deterministic_downscaling_2}, that out of the two possible 
cases when downscaling (over-approximation and under-approximation), we would always under-approximate the value. 
By restricting ourselves to this case, we would cut down significantly on the circuit needed to perform the error compensation by omitting any comparators and control logic. And our tests show that this has no noticeable effect on the accuracy of the operation. 
We would know that the erroneous bit in our downscaled bit-stream is 
always the first 0 we encounter in our thermometer encoded bit-stream; and to compensate for this error,
we would always have to invert this 0 to 1, a certain number of times during our clock division operation.

We know how many bits we need to invert, but we now face the challenge of determining which position of the bits to invert. The erroneous bit is repeated \textit{q} times, and there are \textit{q} candidate positions to perform the \textit{e} (i.e., error magnitude) bit flips. It turns out that we can decide these positions in a deterministic fashion by performing another multiply operation.

\subsection {Multiplication within an Operation} \label{ssec:DoubleMultiplication}

The bit flips need to occur in the right proportion. In other words,
each bit flip of the first operand should be distributed
equally among all the bits of the second operand. 

Take the example discussed earlier in \cref{ssec:error_compensation}. $A'$ (the downscaled bit stream of $A$) has an error of $1$, 
or in other words, one of the 0s should be flipped to 1 and this needs to be distributed among the bits of B'. 
Since $B'$ represents $\frac{3}{4}$, 
it makes sense for that bit flip to align with a $1$ in the bit stream for $B'$. 
On the same line $B'$ has an error of $3$, and needs to be distributed among the bits of $A'$. 
With $A$ representing $\frac{1}{4}$, in order to distribute those bit flips uniformly, 
we would align only one of those bit flips with a 1 in the bit stream of A, and the remaining with 0s.

Trying to figure this distribution out off the top of one's head is easy, but we need a way to compute this deterministically using digital logic. 
We can do this with a multiply operation. In our example for $A'$, we can compute {\it $3 \times
\frac{1}{4} = \frac{3}{4} \approx 1 $}. In fact, we can do so with another {\em unary multiply operation}.

In the example shown in ~\cref{fig:case3b}, based on the error, we would
need to invert one bit of $A$ and three bits of $B$. Since we are always
under-approximating our input operands (and consequently, the result), we will always be changing 0's to 1's. 
For a bit stream $X$, let $Error(X)$ be the number of bits we need to invert, and $Inv(X)$ be the number of inverted bits that need to align with a 1 from the other operand.  The error compensation is illustrated below.

\begin{equation} \label{eqn:1}
\centering
\begin{split}
   \text{Inv}(A') & = \text{Error} \, A' \times B' \\
& = 1 \times \frac{3}{4} = 1
\end{split}
\end{equation}
\begin{equation} \label{eqn:2}
\centering
\begin{split}
   \text{Inv}(B') & = \text{Error} \, B' \times A' \\
& = 3 \times \frac{1}{4} = 1.
\end{split}
\end{equation}

Now that we know where to align those bit flips, we perform the multiply operation with error compensation (bit-flips) as shown below.

\begin{figure}[h]
\centering
\includegraphics{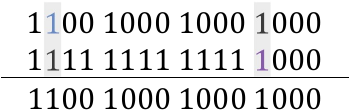}
\label{fig:case3c}
\vspace{-1 em}
\end{figure}

The result of this operation is an output bit-stream corresponding to the value
$\frac{5}{16}$. This is our desired result, as $\frac{5}{16} \times \frac{15}{16} =
\frac{4.6875}{16}$ which is optimally represented as $\frac{5}{16}$.

It is important to note that even with error compensation, it is still possible for our
output bit-stream to not be an optimal approximation. This is because the multiply
operations performed in \cref{eqn:1} and \cref{eqn:2} are carried out with the downscaled values of
our original input operands $A$ and $B$, and hence, there is an approximation involved.
However, {\it the error is bounded to be \emph{at most 2 bits}, regardless of the length of the
input operands}. This is because, in \cref{eqn:1} and \cref{eqn:2}, $A'$ and $B'$ can
have an error of at most 1 bit (out of $2^{n/2}$ bits) from the original values of $A$
and $B$. Consequently, the values
obtained for Inv($A'$) and Inv($B'$) can also differ by at most 1 from their optimal
values. Stated differently, when performing the inversion, the maximum error that can be
introduced is two bits (one for A, and one for B). This would translate to a maximum
error of only two bits at the output, irrespective of the bit-precision of the inputs. 

In fact, in our tests where we did an exhaustive simulation of all possible operations of operands of different bits of precision; we found that less than 0.01\% of cases result in an error of 2 bits, as shown in \cref{ssec:Multiplication_Results}.

It is possible to eliminate this minor error as explained in \cref{sec:Proof}, but the logic involved to do so does significantly impact the overall circuit area. And considering the low error to begin with, as well as fault-tolerant nature of the applications that stochastic/unary computation is usually employed in, we believe the increased gate cost is not justified. 

The method we propose shares a lot of similarities with multiplication using partial products in the binary domain. 
We divide the bits of the operands ($A$ and $B$) into higher-order ($A_h$ and $B_H$), and lower-order ($A_L$ and $B_L$) bits.
The higher-order bits constitute the down-scaled input operands, while the lower order bits represent the error.
The error is compensated by inverting bits, and where we invert those bits is determined by two multiplications: $A_L \times B_H$, and $B_L \times A_H$ . We use these results to correct for the error in our main multiplication of our downscaled operands ($A_H \times B_H$). The one divergence is that we are not performing the multiplication of the lower-order bits, i.e., $A_L \times B_L$. This aspect was initially part of our design, and in fact, eliminates the minute error (max bound of 2 bits) discussed earlier. But this minute improvement in accuracy is accompanied by a $\approx$ 30\% increase in gate cost. We believe that the trade-off is not worth it.

In our example, we have illustrated how to perform multiplication using 16-bit length streams, which conveniently has a square root. However, the proposed technique can still be applied to bit streams of all lengths that are powers of 2, with the caveat that in the cases where we are down-scaling to a length that is not the square-root, there would be an imbalance in the pipelining due to the difference in latencies of the two stages of the operation. 

\section {Mathematical Proof}\label{sec:Proof}

Let us consider two operands \emph{A} and \emph{B}, and let the result of the operation  \emph{$A \times B$}  be \emph{C}. The two operands are both represented as fractions with the denominator being $n$. This corresponds to $n$ bits in the bit stream for the input operands. Consequently, the output bit steam will have a length of $n^2$.

\begin{equation} \label{eqn:3}
\centering
\frac{C}{n^2} = \frac{A}{n} \times \frac{B}{n}    
\end{equation}

We want the output bit-stream to also be $n-bits$. We can rewrite the above equation as:

\begin{equation} \label{eqn:4}
\centering
\frac{C}{n} = n \cdot \left( \frac{A}{n} \times \frac{B}{n} \right)    
\end{equation}

\begin{equation} \label{eqn:5}
\centering
\frac{C}{n} = \left( \frac{A}{\sqrt{n}} \times \frac{B}{\sqrt{n}} \right)    
\end{equation}

$\frac{A}{\sqrt{n}}$ and $\frac{B}{\sqrt{n}}$ are not always integers, let's represent them in terms of quotients and remainders.

\begin{equation} \label{eqn:6}
\centering
\frac{C}{n} = \left( Quotient \left[ \frac{A}{\sqrt{n}} \right] + Remainder \left[ \frac{A}{\sqrt{n}} \right] \right)   \cdot \left( Quo \left[ \frac{B}{\sqrt{n}} \right] + Rem \left[ \frac{B}{\sqrt{n}} \right] \right)
\end{equation}

Expanding the double brackets, we get

\begin{equation} \label{eqn:7}
\centering
\begin{split}
\frac{C}{n} & = \left( Quo \left[ \frac{A}{\sqrt{n}} \right] \cdot Quo \left[ \frac{B}{\sqrt{n}}  \right] \right) + \left( Rem \left[ \frac{A}{\sqrt{n}} \right] \cdot Quo \left[ \frac{B}{\sqrt{n}} \right] \right) \\
& + \left( Quo \left[ \frac{A}{\sqrt{n}} \right] \cdot Rem \left[ \frac{B}{\sqrt{n}}  \right] \right) + \left( Rem \left[ \frac{A}{\sqrt{n}} \right] \cdot Rem \left[ \frac{B}{\sqrt{n}} \right] \right)    
\end{split}
\end{equation}

In our design, the quotients correspond to $A'$ and $B'$, while the remainders are the Error associated with $A'$ and $B'$ respectively.
The terms in \cref{eqn:7} also correspond to different parts of the operation.

\begin{itemize}
    \item {The first term corresponds to the main multiply operation with the downscaled inputs $A'$ and $B'$}
    \item {The second term corresponds to how many bits of $A'$ that we should invert, i.e., Inv(A') in \cref{eqn:1}}
    \item {The third term corresponds to how many bits of $B'$ that we should invert, i.e., Inv(B') in \cref{eqn:2}}
    \item {The fourth term is not considered in our design, but it can be incorporated to completely eliminate any error with respect to the optimal approximation.}
\end{itemize}

Another way of looking at this is that our initial result of multiplying the downscaled inputs $A'$ and $B'$ (i.e., the first therm in \cref{eqn:7} will always be an under-approximation since the inputs were under-approximated. And we correct that under-approximation by inverting/flipping ``0'' bits to ``1'' bits. The matter of ``how many'' and ``where'' to perform these bit flips is computed by the second and third terms in \cref{eqn:7}.

\section {Hardware Implementation}\label{sec:HardwareImplementation}

\begin{figure}
\centering
\includegraphics{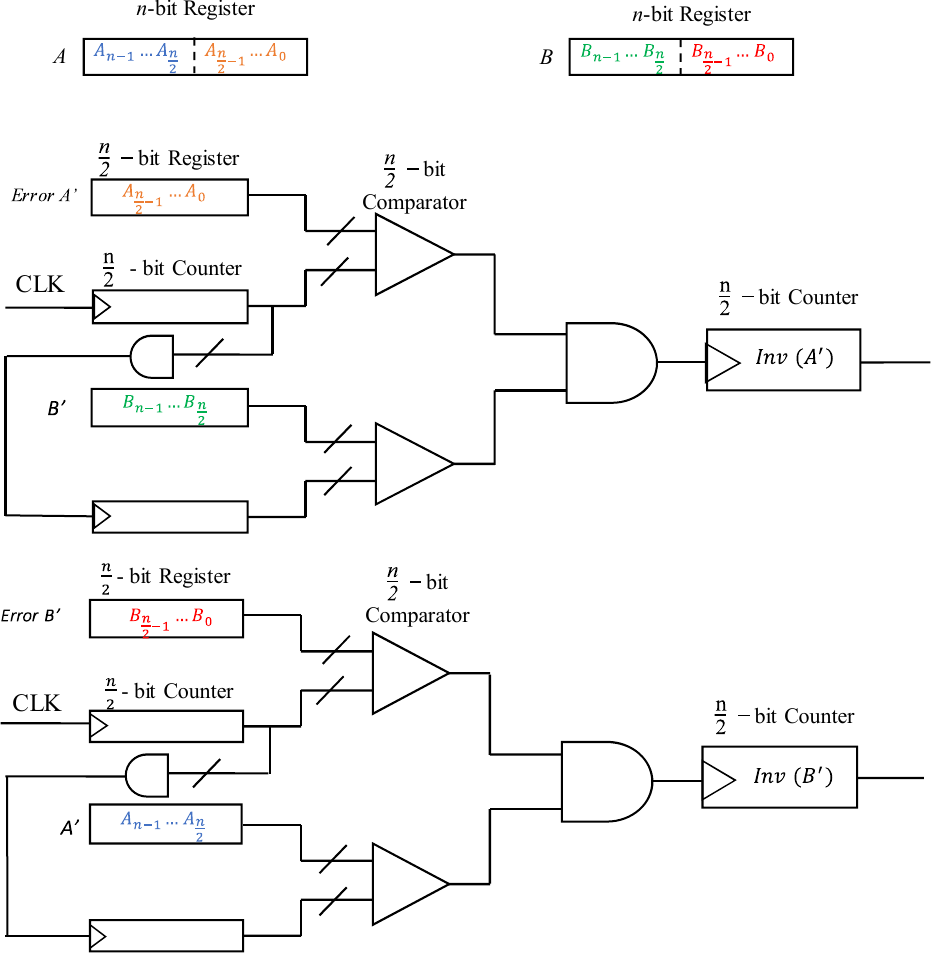}
\caption{}
\label{fig:final_circuit_a}
\end{figure}

\begin{figure}
\centering
\includegraphics{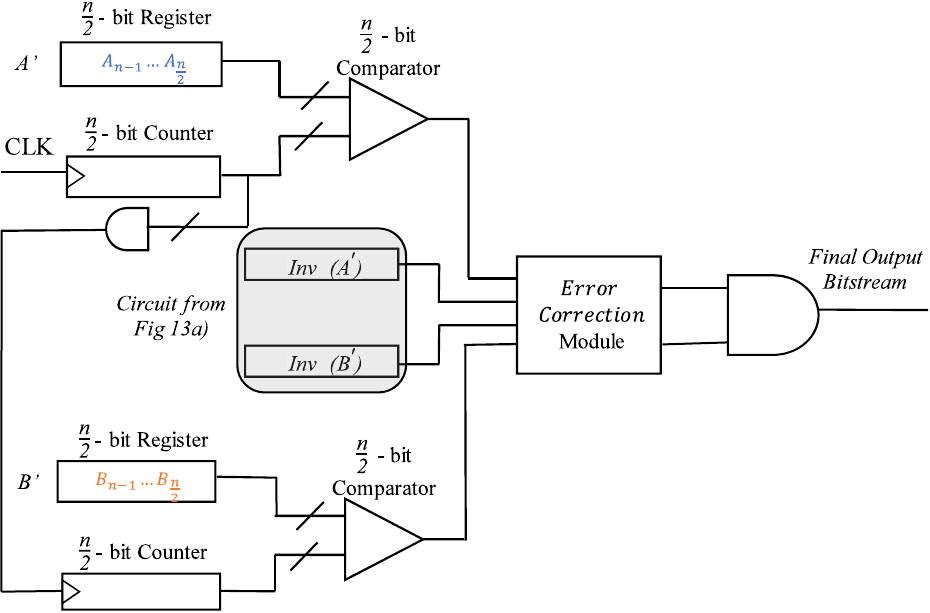}
\caption{}
\label{fig:final_circuit_b}
\end{figure}

The complete circuit for our method is shown in \cref{fig:final_circuit_a} and \cref{fig:final_circuit_b}. 
By 
downscaling the input length to the square root of its original value, the binary values
of $A$ and $B$ can be partitioned in half, as shown in the figure. The higher-order bits represent our downscaled operands, while the lower-order bits represent the error. 

\cref{fig:final_circuit_a} represents the first stage of our operation, responsible for computing \cref{eqn:1} and \cref{eqn:2}. 
It employs two deterministic unary multiplier circuits, each with two unary bit stream generators. 
The generated bit streams are fed to an AND gate which performs the multiplication, and the result is 
accumulated using a counter.

The results from \cref{fig:final_circuit_a} are used in \cref{fig:final_circuit_b},
which carries out the second stage of the operation, i.e., the main multiply operation. 
\cref{fig:final_circuit_b} features two unary bit stream generators for our downscaled input operands, 
which are then fed to an Error Compensation Module that performs the bit flips, and is then fed to a AND gate.

The Error Compensation Module consists of logic that computes
the input to the selector line for two multiplexers: one that chooses between
$A$ and NOT($A$), and the
other between $B$ and NOT($B$). The outputs of these multipliexers serve as the final input to
an AND. The output of the AND gate is accumulated into a $n$-bit counter
and would be the final result of our multiply operation.

Initially, we set out with the goal to deterministically compute the multiplication of two
$2^n$ length input bit streams. We then downscale them to $2^{n/2}$ length input bit streams, to
produce an output bit-stream of length $2^{n}$. This introduces errors in the resultant bit stream since we are dealing with approximations, and we want the optimal approximation for our result.  This error can be deterministically quantified (and compensated) by two other multiply operations, which also involve $2^{n/2}$ bit-stream. These operations can happen in parallel. Therefore, to produce the desired output bit-stream of length $2^{n}$ bits, the latency is $2^{n} +
2^{n} = 2^{n+1}$.
However, there is another optimization that can be implemented. The {\em two stages} of this operation, i.e., determining the error and multiplication with error compensation, can be pipelined to maintain the throughput of one multiply operation every $2^n$ bits.  

\section {Simulation and Results} \label{sec:Results}

We first evaluated our approach with an exhaustive simulation of multiplication of all ${n-bit}$ operands. We compare it to prior stochastic
implementations which rely on pseudo-random or quasi-random generation of bit-streams,
such as LFSRs, Sobol and Halton sequences~\cite{sobol}. We can consider the Sobol sequence
implementation to be representative of all approaches that rely on quasirandom sequences
called low-discrepancy sequences, as they all showcase similar accuracy and area cost.
We then evaluated the different stochastic approaches in other arithmetic functions, and Matrix dot-product to see how they fare in a practical application, 
as it is an integral aspect of machine learning models.

\subsection{Multiplication} \label{ssec:Multiplication_Results}

Table \ref{tab:AccuracyComparison} shows the Mean Absolute Error (MAE) Percentage and Gate Cost of
various implementations for the stochastic multiplication of two inputs. We set the area of the Sobol-Sequences approach as our reference for comparisons.
It is worth mentioning that Sobol-sequences isn't one specific sequence, but rather any sequence in base 2 that satisfies the low-discrepancy/uniformity properties demanded.
For our tests, the two Sobol sequences that had the lowest gate cost, were chosen to generate the bitstreams for the two corresponding operands.

The output bit streams were computed for {\it all possible values} of input operands of
length  $2^n$, and the output was also observed for $2^n$ cycles.  The absolute error was measured against the ideal approximation, and not the full-precision output.
Mathematically, we would need to observe the output from $2^{2n}$ cycles to obtain no error
at all. And in the cases of both Sobol sequences (and other low-discrepancy sequences), and the deterministic approach, 
the error does converge to 0 if the output bitstream were to be generated for $2^{2n}$ cycles.

\begin{table}[h]

\centering
\resizebox{\linewidth}{!}{%
\begin{tabular}{|>{\hspace{0pt}}m{0.123\linewidth}|>{\hspace{0pt}}m{0.138\linewidth}|>{\hspace{0pt}}m{0.138\linewidth}|>{\hspace{0pt}}m{0.138\linewidth}|>{\hspace{0pt}}m{0.123\linewidth}|>{\hspace{0pt}}m{0.138\linewidth}|>{\hspace{0pt}}m{0.138\linewidth}|} 
\hline
\multirow{2}{0.123\linewidth}{Bitstream \\Length} & \multicolumn{2}{>{\centering\arraybackslash}m{0.276\linewidth}|}{LFSR} & \multicolumn{2}{>{\centering\arraybackslash}m{0.261\linewidth}|}{Sobol Sequence} & \multicolumn{2}{>{\centering\arraybackslash}m{0.276\linewidth}|}{Our Approach} \\ 
\cline{2-7}
                                                                                                      & MAE       & Gate Cost                                     & MAE       & Gate Cost                                               & MAE       & Gate Cost                                              \\ 
\hline
$2^4$                                                                                                 & $8.84\%$ & $53.29\%$                                     & $5.93\%$ & $100\%$                                                 & $0.93\%$ & $68.73\%$                                              \\ 
\hline
$2^6$                                                                                                 & $5.35\%$ & $47.28\%$                                     & $1.66\%$  & $100\%$                                                 & $0.34\%$ & $63.16\%$                                               \\ 
\hline
$2^{8}$                                                                                               & $0.96\%$  & $43.05\%$                                     & $0.4\%$  & $100\%$                                                 & $0.16\%$  & $57.73\%$                                              \\
\hline
\end{tabular}
}
\caption{}
\label{tab:AccuracyComparison}
\end{table}

\begin{figure}[!h]
\centering
\includegraphics[width=\linewidth]{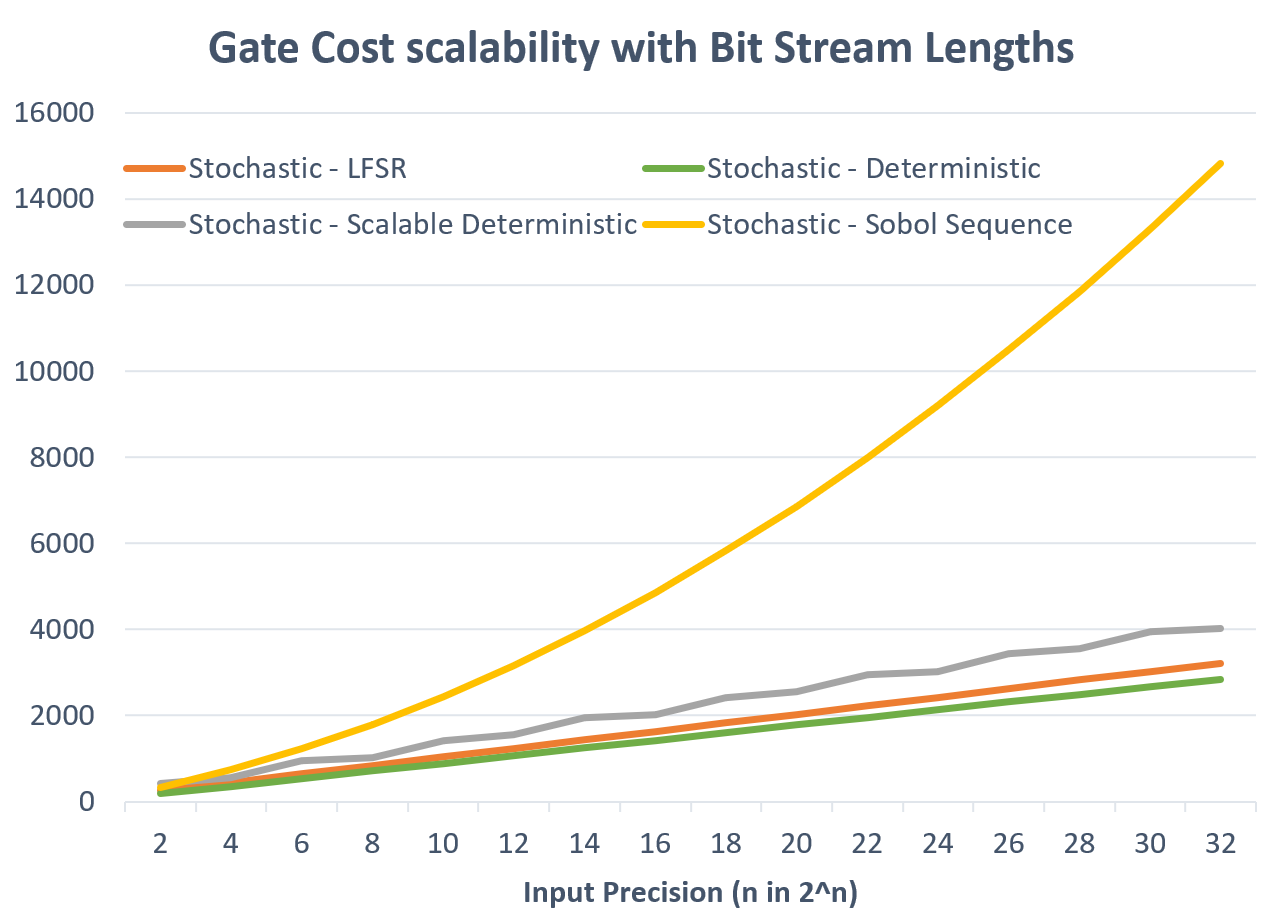}
\caption{}
\label{fig:GateCost1}
\vspace{-1em}
\end{figure}

Our approach offers significant improvements in accuracy over both conventional stochastic implementations
that use LFSRs and other low-discrepancy sequences. Although our approach does demand a slightly higher gate cost over conventional LFSRs, as shown in ~\cref{fig:GateCost1},
the increase in area is minor.
On the other hand, low-discrepancy sequences
such as the Sobol
sequence is accompanied by a large increase in area cost. The gate cost for such
implementations scale quadratically as the precision of the input operands increase, as evident in \cref{fig:GateCost1}. 
This is due to the fact that such low-discrepancy sequences incorporate a Directional Vector
Array in their circuit, whose gate cost scale by a factor of $n^2$ \cite{sobolckt}.

One benefit that low-discrepancy sequences do offer over deterministic approaches is better $progressive$ accuracy, as shown in \cref{tab:Progressive Accuracy}. 
This is due to the innate nature of the distribution of the points in low-discrepancy sequences, and also because deterministic approaches are designed with the assumption that the output is only expected to be read after a certain number of cycles.
However, we argue that this is irrelevant as the desired precision of the output
is predetermined in the design phase of an application, and remains constant.

\begin{table}[]
\centering
\begin{tabular}{|c|cccc|}
\hline
\multirow{2}{*}{\begin{tabular}[c]{@{}c@{}}Observed Output Bitstream Length \\ (Bits)\end{tabular}} & \multicolumn{4}{c|}{Mean Absolute Error (\%)}                                                       \\ \cline{2-5} 
                                                                                                    & \multicolumn{1}{c|}{LFSR} & \multicolumn{1}{c|}{Sobol} & \multicolumn{1}{c|}{Halton} & Our Approach \\ \hline
10                                                                                                  & \multicolumn{1}{c|}{31.53} & \multicolumn{1}{c|}{18.96}  & \multicolumn{1}{c|}{19.74}   & 23.67                             \\ \hline
11                                                                                                  & \multicolumn{1}{c|}{28.36} & \multicolumn{1}{c|}{16.34}  & \multicolumn{1}{c|}{17.21}   & 18.45                              \\ \hline
12                                                                                                  & \multicolumn{1}{c|}{24.96} & \multicolumn{1}{c|}{13.74}  & \multicolumn{1}{c|}{14.89}   & 12.32                              \\ \hline
13                                                                                                  & \multicolumn{1}{c|}{22.87} & \multicolumn{1}{c|}{10.8}  & \multicolumn{1}{c|}{10.33}   & 8.61                              \\ \hline
14                                                                                                  & \multicolumn{1}{c|}{18.6} & \multicolumn{1}{c|}{7.36}  & \multicolumn{1}{c|}{8.1}   & 3.78                              \\ \hline
15                                                                                                  & \multicolumn{1}{c|}{13.5} & \multicolumn{1}{c|}{6.84}  & \multicolumn{1}{c|}{7.47}   & 1.52                              \\ \hline
16                                                                                                  & \multicolumn{1}{c|}{8.84} & \multicolumn{1}{c|}{5.93}  & \multicolumn{1}{c|}{6.13}   & 0.93                              \\ \hline
\end{tabular}
\caption{}
\label{tab:Progressive Accuracy}
\end{table}

\subsection{Arithmetic Functions} \label{ssec:Arithemetic_Functions_Results}
The proposed method can be applied to
many stochastic operations. \cite{parhi1, parhi2} demonstrates how to perform operations such as exponent, sin, log
in the stochastic domain using NAND gates to implement the Maclaurin series expansion of these functions. For these tests, 
we settled on bit-streams of length $2^8$ bits, as it provides a good balance of accuracy, precision and latency.
We do make some minor adjustments such that the coefficients in polynomial are approximated such that 
the denominator's precision is $\frac{1}{2^8}$. The increase in error due to this change is offset by increasing the degree
of the polynomial, which translates to more levels of logic. As with multiplication, the tests were exhaustive, covering all possible values of $2^n$-bit operands.

The Mean Absolute Error (MAE) and gate cost are shown in \cref{tab:AccuracyComparison2}. The general trend continues; our technique offers better
accuracy than the state-of-the-art Sobol sequences, while offering significant reductions in area. In some cases, the gate cost of 
Sobol sequences is over twice our proposed circuit. And the gap only widens as we scale the length of the bit-streams.

\begin{table}
\captionsetup{justification=centering}

\centering
\resizebox{\linewidth}{!}{%
\begin{tabular}{|>{\hspace{0pt}}m{0.156\linewidth}|>{\hspace{0pt}}m{0.127\linewidth}|>{\hspace{0pt}}m{0.135\linewidth}|>{\hspace{0pt}}m{0.127\linewidth}|>{\hspace{0pt}}m{0.121\linewidth}|>{\hspace{0pt}}m{0.135\linewidth}|>{\hspace{0pt}}m{0.135\linewidth}|} 
\hline
\multirow{2}{0.154\linewidth}{\hspace{0pt}Operation} & \multicolumn{2}{>{\hspace{0pt}}m{0.262\linewidth}|}{LFSR} & \multicolumn{2}{>{\hspace{0pt}}m{0.248\linewidth}|}{Sobol Sequence} & \multicolumn{2}{>{\hspace{0pt}}m{0.27\linewidth}|}{Our Approach}  \\ 
\cline{2-7}
                                                     & MAE      & Gate Cost                                      & MAE      & Gate Cost                                                & MAE       & Gate Cost                                             \\ 
\hline
$e^{-x}$                                             & $7.2\%$ & $44.27\%$                                      & $3.3\%$ & $100\%$                                                  & $1.6\%$ & $53.32\%$                                             \\ 
\hline
$sin$ $x$                                             & $7.9\%$ & $48.67\%$                                      & $3.1\%$ & $100\%$                                                  & $1.5\%$  & $57.20\%$                                             \\ 
\hline
$log (1+x)$                                          & $6.7\%$ & $45.31\%$                                      & $3.6\%$ & $100\%$                                                  & $1.9\%$  & $48.02\%$                                             \\ 
\hline
$sigmoid$ $x$                                         & $8.4\%$ & $52.76\%$                                      & $3.2\%$ & $100\%$                                                  & $1.4\%$  & $60.61\%$                                             \\
\hline
\end{tabular}
}
\caption{}
\label{tab:AccuracyComparison2}
\end{table}

\begin{figure}[h]
\centering
\includegraphics[width=\linewidth]{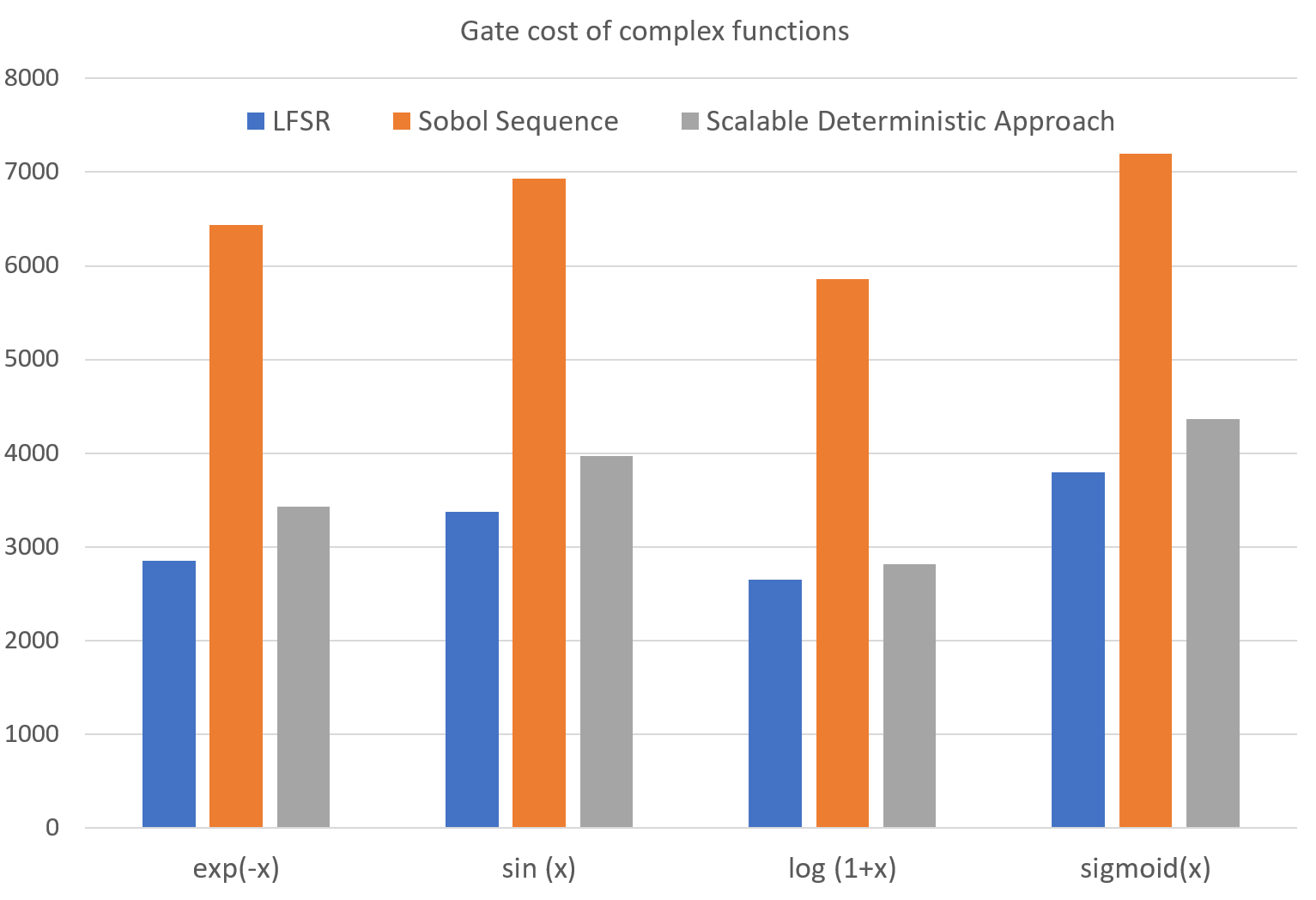}
\caption{}
\label{fig:GateCost2}
\vspace{-1em}
\end{figure}

\subsection{Matrix Multiplication} \label{ssec:Matrix_Multiplication}

Error tolerance, combined with many low precision operations, make ML models an ideal candidate for stochastic computing.
\cite{Liu-survey} is comprehensive survey of different neural networks that incorporate the technique. 

The three key computations performed in a ML model are: matrix/vector multiplication, accumulation (i.e., addition), and the activation function. Our focus in this paper is matrix dot-product multiplication. Although several designs \cite{Hayes-Matrix}\cite{Hayes-Hybrid} have been proposed to perform accumulation in the stochastic domain \cite{Liu-survey}, accumulation in the traditional binary domain generally works better. This is because stochastic logic is limited to the range [0, 1] so accumulation requires scaling. Activation functions are heavily reliant on the design of the ML model. If one wishes to compute the activation function in the stochastic domain, in most cases one can do so via arithmetic functions such as \emph{Btanh} and \emph{Sigmoid} \cite{activation-btanh}\cite{activation-sigmoid}. 

\begin{figure}[h]
\centering
\includegraphics[width=\linewidth]{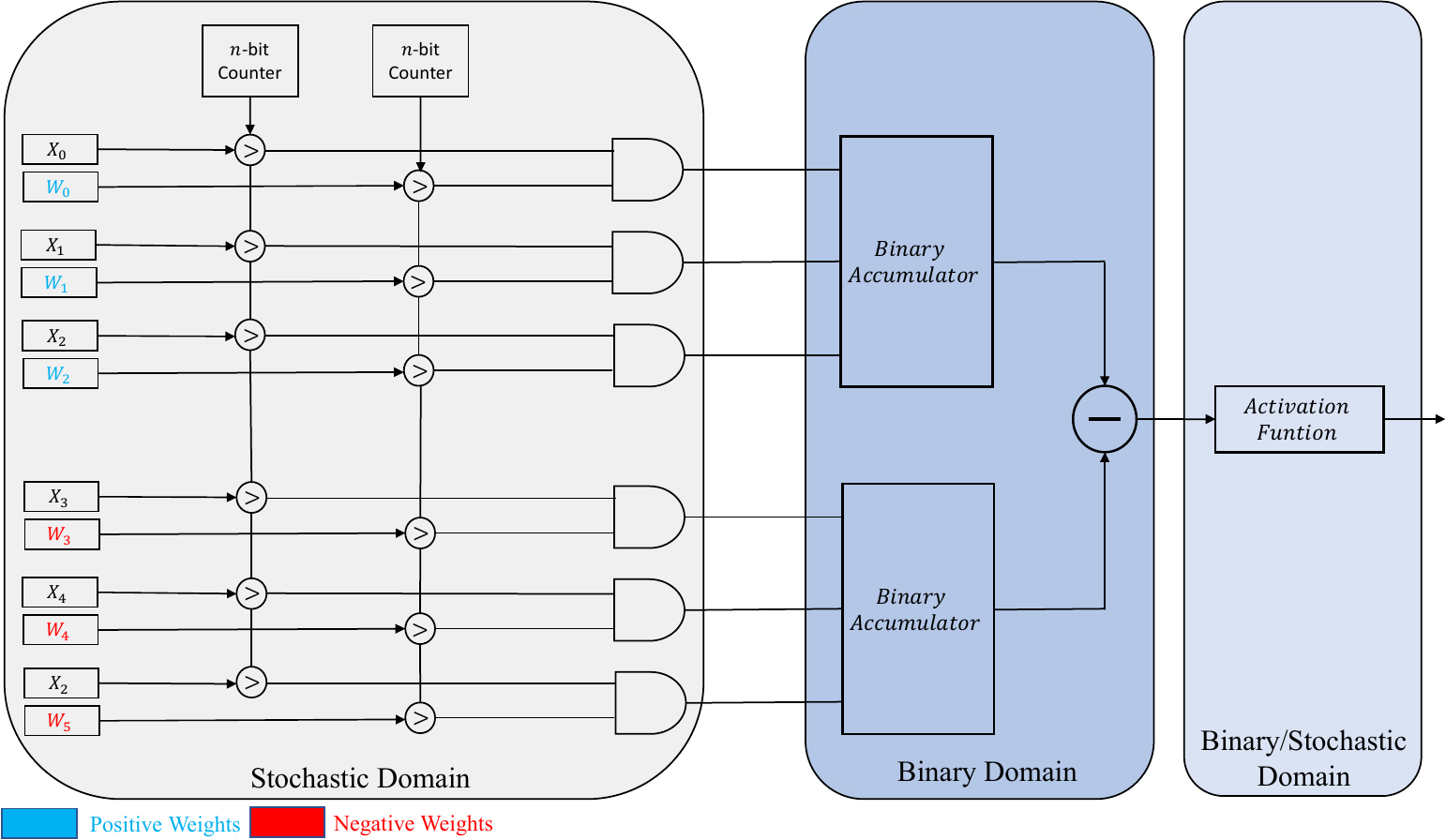}
\caption{}
\label{fig:Stochastic-Neuron}
\end{figure}

For neural network computation, we have to address the issue of negative weights. Although stochastic computing can support negative values within the range [-1,1] by using the \emph{bipolar} representation, that approach increases latency and gate cost due to additional processing. Furthermore, it does not scale well. Since binary adders are more efficient than stochastic ones, we implement positive and negative weights separately, and we perform accumulation in the binary domain. 
\cref{fig:Stochastic-Neuron} demonstrates how a neuron can be modeled and implemented with positive and negative weights. The same counter can be used to generate the bit-streams for all elements. However, each input operand bit-stream requires exclusive access to a comparator circuit. The designer has the choice of how many comparator circuits they want to incorporate, based on the priority of latency or area for that particular design.

We simulated the dot product with two matrices, A and B, of sizes [2048 2048] and [2048 128], respectively. The elements were initialized to random $n$-bit values. The tests were run for 100 trials, and the results were averaged across all trials. \cref{tab:Matrix-MAE} shows the mean absolute error of all the elements in the product matrix $C = A \cdot B$. The same trend observed in \cref{ssec:Multiplication_Results} continues, and in fact, the gap widens. This can be attributed to the fact that, unlike \cref{tab:AccuracyComparison}, this was not an exhaustive simulation across ALL $n$-bit values, but rather, a more realistic scenario with operands initialized to random values of $n$-bit precision.

\begin{table}[]
\centering
\begin{tabular}{|c|cccc|}
\hline
\multirow{2}{*}{\begin{tabular}[c]{@{}c@{}}Input/Output \\ Bit-Stream Length\end{tabular}} & \multicolumn{4}{c|}{Mean Absolute Error (\%)}                                                                            \\ \cline{2-5} 
                                                                                           & \multicolumn{1}{c|}{LFSR} & \multicolumn{1}{c|}{Sobol} & \multicolumn{1}{c|}{Halton} & \multicolumn{1}{c|}{Our Approach} \\ \hline
$2^4$                                                                                          & \multicolumn{1}{c|}{10.47} & \multicolumn{1}{c|}{6.44}  & \multicolumn{1}{c|}{7.32}   & 0.87                              \\ \hline
$2^6$                                                                                          & \multicolumn{1}{c|}{6.83} & \multicolumn{1}{c|}{2.16}   & \multicolumn{1}{c|}{2.85}    & 0.31                               \\ \hline
$2^8$                                                                                          & \multicolumn{1}{c|}{3.86}  & \multicolumn{1}{c|}{1.59}  & \multicolumn{1}{c|}{1.63}   & 0.26                              \\ \hline
\end{tabular}
\caption{}
\label{tab:Matrix-MAE}
\end{table}

The design presented in \cite{Sobol-neural} incorporates stochastic computing and low-discrepancy Sobol sequences in the first convolution layer of the LeNet-5 neural network. We reconstructed the test environment and substituted the stochastic operations with deterministic unary operations. As shown in \cref{tab:Lenet-test}, using a deterministic unary approach achieves better classification rates than other alternative random/quasirandom number generation schemes, at a much lower area cost.

\begin{table}[]
\centering
\begin{tabular}{|c|c|}
\hline
Design            & Misclassification Rate for $2^4$ operating cycles \\ \hline
Conventional LFSR & 1.08\%                                            \\ \hline
Sobol sequences   & 0.84\%                                            \\ \hline
Our approach      & 0.79\%                                            \\ \hline
\end{tabular}
\caption{}
\label{tab:Lenet-test}
\end{table}

We are not advocating for a specific ML model or architecture. Instead, the goal of our design is to offer a flexible and scalable method for performing multiplication in the stochastic domain. The deterministic approach is designed to provide a cost-effective (in terms of gates) and adaptable framework that is well-suited for fault-tolerant and low-precision applications. While stochastic computing offers area savings over conventional binary circuits for higher-precision operations, the associated latency proves to be limiting and cannot surpass the balance of area-latency offered by traditional binary computing.

\section {Conclusion} \label{sec:conclusion}

Recent work has demonstrated that randomness is not a requirement for ``stochastic''
computing. The deterministic approach in~\cite{devon2} mitigates most
of the drawbacks typically associated with the paradigm. 
However, the method in these papers does not allow for 
graceful approximations when constant bit-stream lengths are required.

In this paper, we presented an approach that builds upon this foundation.  By
deterministically downscaling the inputs and compensating for approximation errors during
the clock division operation, we demonstrate that it is possible to produce accurate
results, while also preserving the bit stream lengths. This makes our approach {\em
composable}, allowing operations to be chained together.  Our simulations show that our
approach can achieve very accurate results, with the maximum error bounded as two bits
for each level of logic, irrespective of the bit stream length. It offers
significant advantages over other stochastic approaches that rely on random or
quasi-random bit streams. And it serves as a viable energy/area efficient alternative to traditional binary computation
in low-precision applications that are fault-tolerant and less latency-sensitive.

\section*{Author Contributions}

YK and MR led discussions on this research. YK conducted the data analyses and wrote the manuscript. YK and MR reviewed the manuscript.

\section*{Funding}

The authors are funded by DARPA grant \#W911NF-18-2-0032.

\bibliographystyle{Frontiers-Vancouver} 
\bibliography{ref}


\section*{Figure captions}
\begin{itemize}
\item Figure 1: Stochastic implementation of common arithmetic operations: (a) Multiplication; (b) Scaled addition.
\item Figure 2: Unary/Thermometer code generator
\item Figure 3: Multiplication - Conventional Stochastic vs Deterministic approach
\item Figure 4: Circuit implementation of clock-division involving two counters connected in series.
\item Figure 5: Circuit to compute the error and which bits to invert 
\item Figure 6: Circuit to perform main multiply operation
\item Figure 7: Relative gate cost for different stochastic implementations of a multiply circuit
\item Figure 8: Relative gate cost for implementation of different arithmetic functions 
\item Figure 9: Structure of hybrid binary-stochastic neuron implementation 
\item Table 1: Mean absolute error and gate cost \% for the multiply operation of various stochastic implementations
\item Table 2: Progressive accuracy comparison between different stochastic approaches
\item Table 3: Mean absolute error and gate cost \% for functions implemented using Maclaurin expansion
\item Table 4: Mean absolute error of different stochastic techniques for matrix dot-product.
\item Table 5: Misclassification rate for LeNeT-5
\end{itemize}

\end{document}